# High Modulation Efficiency and Large Bandwidth Thin-Film Lithium Niobate Modulator for Visible Light

Chijun Li[1], Bin Chen[2], Ziliang Ruan[2], Pengxin Chen[1], Kaixuan Chen[1], Changjian Guo[1, †], and Liu Liu[2, †]

[1] Guangdong Provincial Key Laboratory of Optical Information Materials and Technology, South China Academy of Advanced Optoelectronics, Sci. Bldg. No. 5, South China Normal University, Higher-Education Mega-Center, Guangzhou 510006, China.

[2] State Key Laboratory for Modern Optical Instrumentation, Centre for Optical and Electromagnetic Research, Zhejiang Provincial Key Laboratory for Sensing Technologies, East Building No. 5, Zijingang Campus, Zhejiang University, Hangzhou 310058, China.

**Abstract:** We experimentally demonstrate a visible light thin-film lithium niobate modulator at 532 nm. The waveguides feature a propagation loss of 2.2 dB/mm while a grating for fiber interface has a coupling loss of 5 dB. Our demonstrated modulator represents a low voltage-length product of 1.1 V·cm and a large bandwidth beyond 30 GHz.

† Correspondence to: C. Guo, changjian.guo@coer-scnu.org; L. Liu, liuliuopt@zju.edu.cn

## 1. Introduction

Underwater communication is extensively used in deep-sea resource exploration, marine ecosystem monitoring, underwater rescue, and information exchange between autonomous underwater vehicles[1]. Traditional acoustic underwater communication systems have low bandwidth, severe multipath delay, and high energy consumption, which cannot meet the demand of the ever-increasing data volume[2]. In addition, radio frequency (RF) communication is severely restricted in the ocean due to the large decay of RF signals in seawater[3,4]. Compared with acoustic and RF underwater communication systems, the underwater wireless optical communication (UWOC) systems present significantly higher transmission bandwidths. UWOC systems based on direct modulation of light-emitting diodes (LEDs) or laser diodes (LDs) have been developed in recent years[5-8]. However, the maximum bandwidth of these systems is only in the order of GHz, which is mostly limited by the bandwidth of its light source. The 3-dB bandwidth of a red-light vertical-cavity surface-emitting laser (VCSEL) can reach 26.3 GHz using two-stage injection locking but at the expense of system complexity and power consumption[9]. Employing external modulation will greatly improve the communication capability of the UWOC system if a high-speed visible light modulator can be developed.

Generally, visible-light integrated devices can be implemented on silicon-nitride, silica, or thin-film lithium niobate (LN) platform due to their low absorption coefficient within the visible spectrum[10,11]. However, silica and silicon nitride have low thermo-optical coefficients and are lack of electro-optical (EO) properties, which makes integrated modulation at visible wavelengths a challenge[12]. LN is one of the most promising materials for electro-optic devices due to its many desirable properties, including a wide transmission spectrum ranging from the visible to the mid-infrared, strong electro-optic effect, stable physical and chemical characteristics, etc., which makes it an excellent candidate for EO modulators at visible light wavelength[13-16].

In this paper, we propose and demonstrate a thin-film LN visible light modulator at 532 nm. The fabricated modulator features an insertion and propagation loss of about 3.05 dB, a $V_\pi$ of about 3.3V, and a bandwidth of more than 30 GHz.

## 2. Design and fabrication

Figure 1 shows the schematic diagram of the modulator. The modulator used a push-pull configuration and was fabricated on a commercial x-cut lithium niobate on silicon (LNOI) wafer (NanoLN, Jinan, China) with a 200-nm thick LN film. The LN waveguide has a ridge height of 100 nm, i.e., half of the thickness of the LN film. A 2-μm thick thermal oxide lies between the thin-film LN and the silicon substrate. A grating coupler (GC) etched in half of the thickness of the LN film was used as the fiber-to-chip interface. A multimode interferometer (MMI) was used as the 3 dB splitter and combiner for the Mach Zehnder interferometer (MZI). The MMI had a relaxed fabrication tolerance as well as a large wavelength bandwidth of 107 nm. Traveling wave (TW) electrodes based on a classic coplanar waveguide (CPW), in which electric and optical waves propagate collinearly, were also adopted in the proposed MZI. The electrodes support a quasi-transverse electromagnetic (quasi-TEM) mode for the electrical signal with low dispersion and therefore can offer a high modulation bandwidth. We will discuss the detailed design and fabrication process of each component in the remaining part of this section.

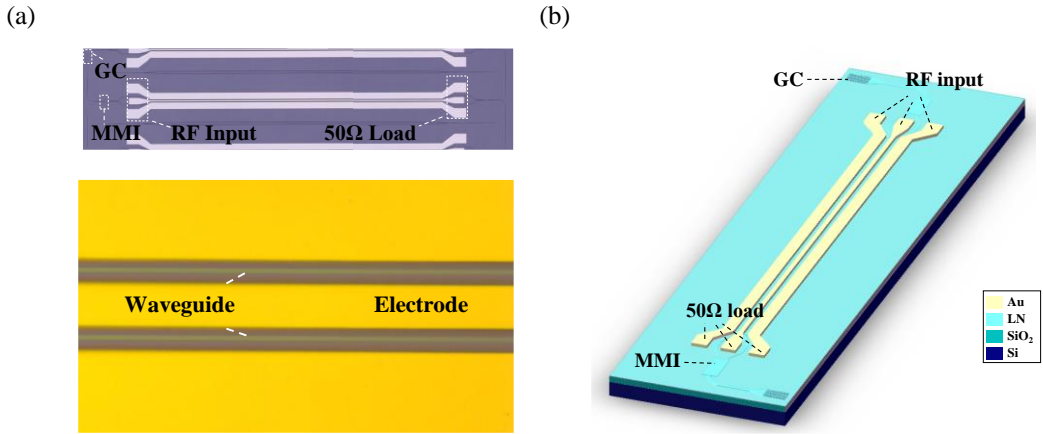

Fig. 1. (a) Optical microscope of the overall sketch and partially enlarged view of the waveguide; (b) 3D structure of the proposed device.

We used GCs as the fiber-to-chip interfaces instead of edge coupling due to the difficulty of fabrication. Fig. 2(a) shows the cross-sectional view for fiber coupling. The fiber was placed on the top of the GC within the symmetry plane, with a tilted angle of -10 degrees with respect to the normal direction of the chip surface. The etching depth $h$ was the same as that of the LN ridge waveguides. The sidewall angle $\alpha$ of all etched LN patterns was set to 60 degrees, a typical value from the current fabrication technology. We simulated the coupling efficiency of the GC as a function of duty cycle ($\rho$) and period ($p$) via numerical scan using a Gaussian beam with a waist diameter of 3.6 μm as the input, calculating the optimal solution. The simulated result of coupling efficiency can also be seen in Fig. 2(b). Accordingly, $p$ and $\rho$ were chosen to be 288 nm and 0.6, respectively, where a maximum coupling efficiency of 33.6% was obtained at 532 nm, corresponding to a device insertion loss of -4.7 dB shown in Fig. 2(c). In addition, a scan-electron microscope (SEM) picture of the fabricated sample is displayed in Fig. 2(d), knowing the device manufacture is relatively well.

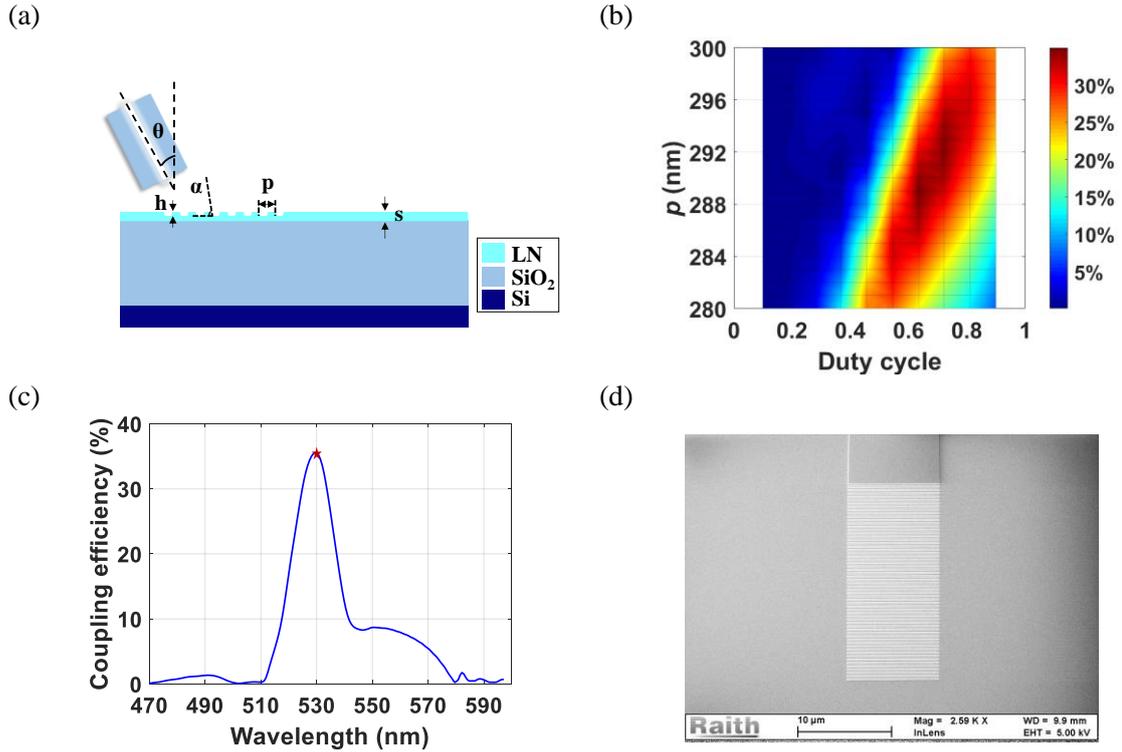

Fig. 2. (a) Cross-section of the grating structure; (b) simulated coupling efficiency with respect to the period and duty cycle; (c) simulated spectra of the designed GC using the optimal parameters ($p$=288, $\rho$=0.6); (d) SEM picture of the fabricated GC.

The 3-dB coupler used in the MZI was realized by a MMI coupler, as shown in Fig. 3(a). The width $W_{MMI}$ and length $L_{MMI}$ of the MMI region were 4 and 36.1 μm, respectively. The width $w$ of the input and output waveguide was 1.5 μm. The separation between the output ports $Y_{MMI}$ was 2.08 μm to ensure low crosstalk between the two output ports. The simulated transmission spectra at the output Port1 and Port 2 is shown in Fig. 3(b). The peak coupling efficiency is 49.6% at 532 nm, corresponding to an insertion loss from the input port to either of the output port of -3.05 dB.

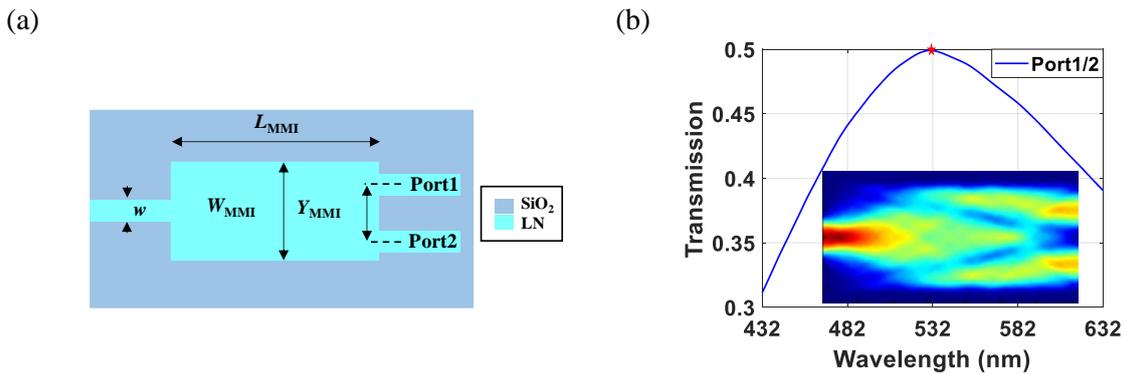

Fig.3. (a) Top view schematic of the 3-dB MMI coupler; (b) calculated two-port transmission of the designed MMI splitter. Inset: simulated light propagation in the designed MMI at 532 nm.

The performance of a traveling wave MZI modulator shown in Fig. 4(a) is mainly determined by the cross-section of the modulation section. Au electrode with a thickness of 1.1 μm was used in the design, which ensures a low microwave loss. The other parameters of the electrodes, i.e., the gap between electrode ($g$), the width of the signal electrode ($w_s$), and the width of the ground electrode ($w_g$) were designed as 5.5, 13.8, 60 μm, respectively. The width $w$ of the waveguide was 1.5 μm. Accordingly, the effective RF refractive index $n_m$, loss $\alpha_m$ and the characteristic impedance $Z_0$ can be obtained via simulation, as shown in Figs. 4(b)-4(d), respectively. From Fig. 4(b), one can see that the refractive index of the driving RF signal ($n_m$) is nearly matched with the group index of the optical mode ($n_o$) shown by the red dashed line. The characteristic impedance ($Z_0$) of the electrode was also designed to be ~50 Ω which is matched with the source and load impedance, as shown in Fig. 4(d). From Fig. 4(c), one can see that the RF loss $\alpha_m$ is less than 10 dB/cm within the range of 120 GHz. The simulated $V_\pi L$ product is 0.96 V·cm, as shown in Fig. 4(e).

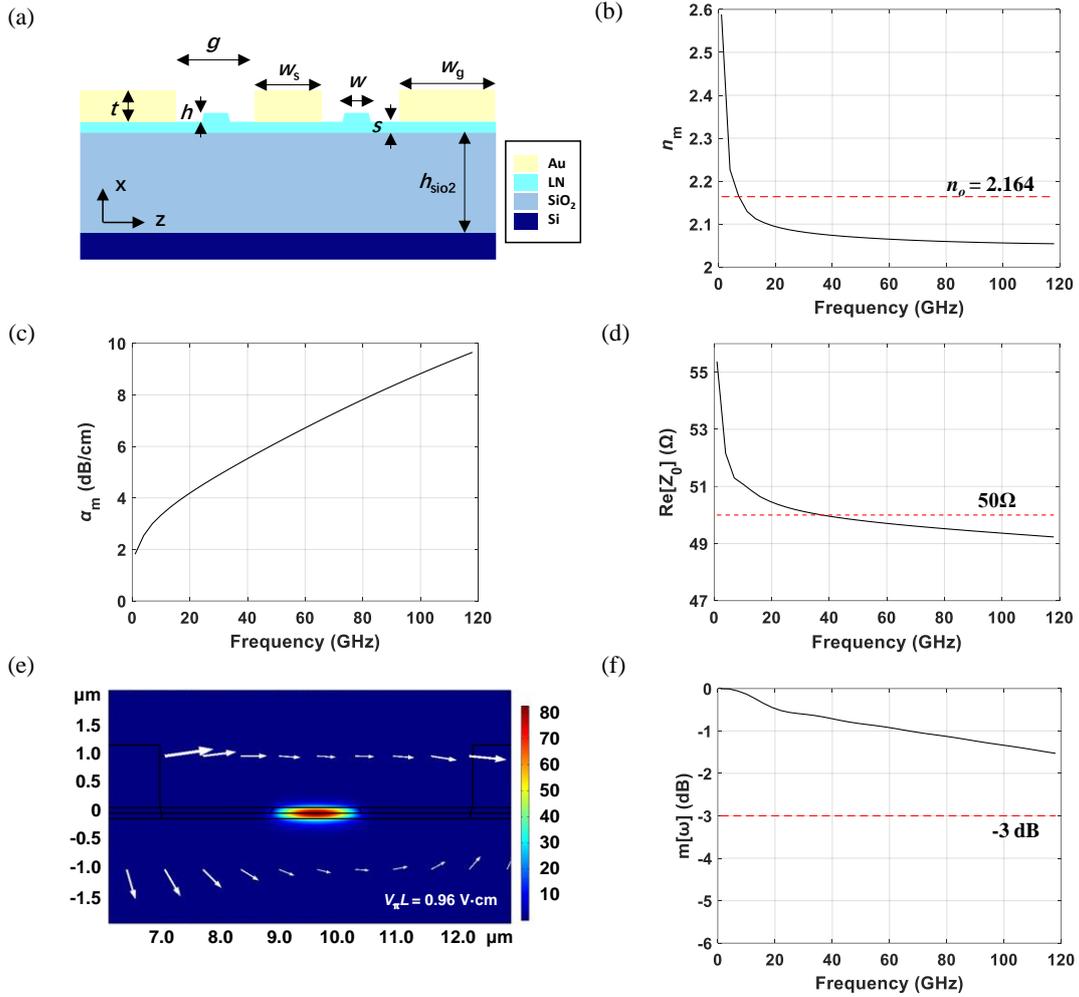

Fig. 4. (a) Cross-section of the visible light modulator; (b) RF effective index $n_m$; (c) RF loss $\alpha_m$; (d) Impedance $Z_0$; (e) Optical and electrical field of the modulator; (f) Calculated modulation response $m[\omega]$ of the designed modulator of 3.3 mm. Here the design is with $g$=5.5 μm, $t$=1.1 μm, $w_s$=13.8 μm, and $w_g$=60 μm.

## 3. Measurements

**Optical loss measurement.** A single-longitudinal-mode 532-nm laser (Changchun Laser Optoelectronics Technology Co., Ltd..) with a spectral linewidth of < 1E-5 nm, and an optical power meter (Newport Model 1918-R) were used for characterization of the insertion loss of the fabricated device. The loss of the GC was first measured using a cascaded GC structure with a back-to-back configuration. One can see from Fig. 5(a) that the measured insertion loss of the GC is 5 dB. We then measured the loss of the MMI by cascading five MMIs. The measured total loss (2 GCs + 5 MMIs) was 29 dB, corresponding to an excess loss for each MMI of about 0.8 dB. We also measured the propagation loss of the LN ridge waveguide by three different lengths. The propagation loss calculated via the slope of the measured loss curve was about 2.2 dB/mm. In addition, it is noted that the main sources of the waveguide loss are scattering due to rough sidewalls, and it is expected to be much more significant at the visible wavelengths considered here since Rayleigh scattering is proportional to $\lambda^{-4}$, where λ is the wavelength of light.

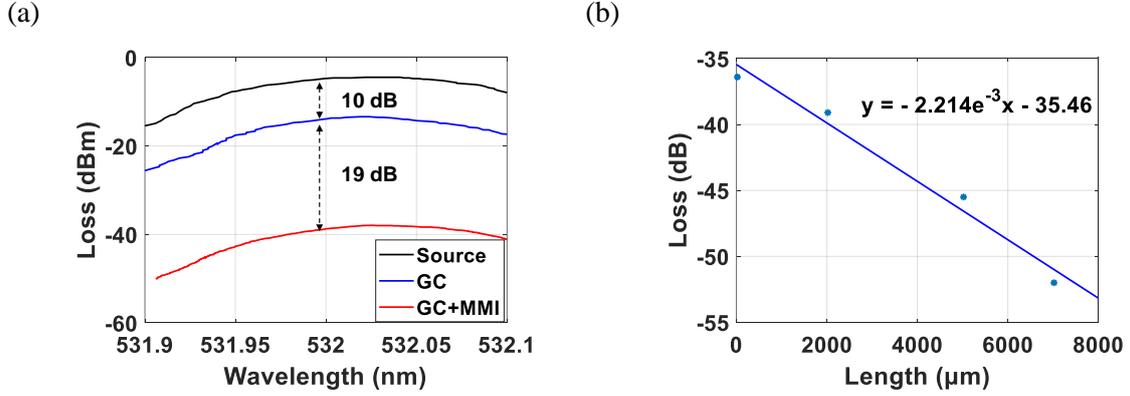

Fig. 5. (a) GC loss and MMI loss; (b) Waveguide propagation loss.

**Electro-optic (EO) performance.** The half-wave voltage ($V_\pi$) of the fabricated modulator was measured using triangular waves. In addition, the devices are driven in a single-drive push-pull configuration, so that applied voltage induces a positive phase shift in one arm and a negative phase shift in the other. Fig. 6(a) shows the measured $V_\pi$ of ~ 3.3 V, which agrees well with the simulated results. The EO response of the fabricated MZI modulators is shown in Fig. 6(b), where the responses of the RF cables, RF probes, photodetector (Newport model 1414, 25GHz), and microwave amplifier (SHF S807) were subtracted during calibration. The curves are noisy due to the high losses as well as low sensitivity of photodetector. Nevertheless, we can see that the measured 3-dB EO bandwidth (S21) is beyond 30 GHz, limited by the bandwidth of the photodetector.

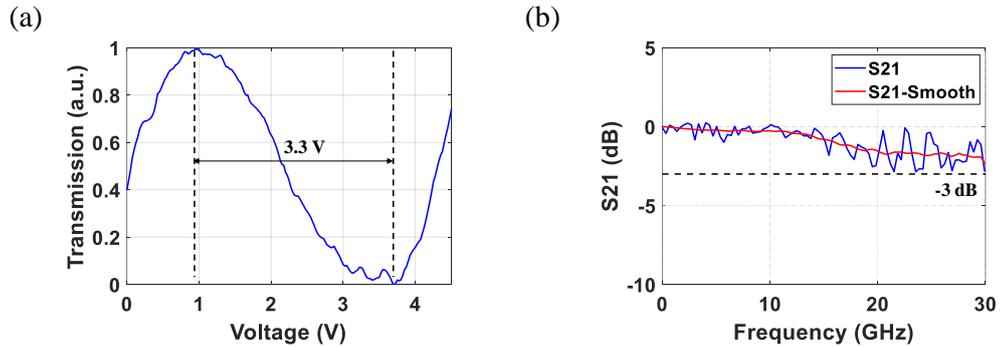

Fig. 6. (a) $V_\pi$ of the modulator; (b) EO S21 of the modulator.

## 4. Conclusion

In summary, we have proposed a high-performance visible light modulator based on the thin-film LN platform. For a modulation length of 3.3 mm, the device can achieve a $V_\pi$ of 3.3 V and an EO bandwidth of beyond 30 GHz. The fabricated modulator has a loss of 18.8 dB. Since the visible light communication technology nowadays is still hindered by the modulation bandwidth, we believe the realization of a large bandwidth external modulator in the visible range will promote the further development of high speed UWOC in various applications.

## Acknowledgments

The authors acknowledge financial support from the Science and Technology Planning Project of Guangdong Province (Grant No. 2019A050510039) and the Guangdong Basic and Applied Basic Research Foundation (Grant No. 2021A1515012215).